\documentclass{iopart}
\usepackage{iopams,amsfonts,amssymb,amsthm}
\usepackage{geometry}
\usepackage{graphicx}
\usepackage{rawfonts}
\usepackage{amssymb}
\usepackage{epstopdf}
\usepackage[usenames,dvipsnames,svgnames]{xcolor}

\begin{document}

\title{Non-Hermitian heat engine with all-quantum-adiabatic-process cycle}
\author{S. Lin and Z. Song\footnote[1]{\textsf{songtc@nankai.edu.cn}}}
\address{School of Physics, Nankai University, Tianjin 300071, China}
\begin{abstract}
As a quantum device, a quantum heat engine (QHE) is described by a Hermitian
Hamiltonian. However, since it is an open system, reservoirs have to be
imposed phenomenologically without any description in the context of quantum
mechanics. A non-Hermitian system is expected to describe an open system
which exchanges energy and particles with external reservoirs.
Correspondingly, such an exchange can be adiabatic in the context of quantum
mechanics. We first propose a non-Hermitian QHE by a concrete simple
two-level system, which is an $S=1/2$ spin in a complex external magnetic
field. The non-Hermitian $\mathcal{PT}$-symmetric Hamiltonian, as a
self-contained one, describes both working medium and reservoirs. A
heat-engine cycle is composed of completely quantum adiabatic processes.
Surprisingly, the heat efficiency is obtained to be the same as that of
Hermitian quantum Otto cycle. A classical analogue of this scheme is also
presented. Our finding paves the way for revealing what the role of a
non-Hermitian Hamiltonian plays in physics.
\end{abstract}

\maketitle





\pagenumbering{arabic}

\section{Introduction}

Non-Hermitian system has received intensive studies due to two reasons: It
can possess full real eigenvalues \cite{Bender1} which is fundamental in the
quantum world. A non-Hermitian potential breaks the probability
conservation, which is supposed to characterize the source or drain of
particles and energy. The connection between the imaginary potentials and
the environment has been investigated in discrete system \cite{JL1}. A
non-Hermitian system could be employed to describe an open system which
exchanges energy and particles with external reservoirs. An alternative way
to deeply understand the exact meaning of a non-Hermitian Hamiltonian in
physics is to find out its function which is similar to some phenomena in
practice. An exemplified example is the invention that an imaginary
potential can be the function of laser or anti-laser \cite{Longhi,ZXZ} when
spectral singularity is reached \cite{Mostafazadeh}.

The topic of the quantum heat engine (QHE) \cite%
{Alicki,Geva,Bender2,Quan2007,Henrich,Abah,YiXXJPA,Altintas} has attracted a
lot of interest since it was first proposed by Scovil and Schultz-Dubois
\cite{Scovil}. A QHE converts heat into useful mechanical work from quantum
working mediums, such as multilevel systems \cite{Quan2005}, harmonic
oscillator systems \cite{Lin,Kosloff,ScullyPRL,Rezek}, spins or coupled
spins \cite{He}, optomechanical systems \cite{ZhangPRL}, relativistic
particles \cite{Munoz} and so on. Many investigations have been carried out
to explore various possible improvement of the heat engine that the
efficiency of a QHE can surpass the efficiency of a classical Carnot heat
engine, and a QHE can extract work from a single heat bath \cite%
{ScullyScience}. Due to the quantum properties of the working medium, or the
effects of the quantum heat bath, some unusual and exotic phenomena have
been manifested by considering and using a squeezed reservoir \cite%
{YiXX2012,Robnagel}, quantum coherence \cite{ScullyScience}, or coupled
spins \cite{ZhangPRA,Thomas,YiXX2013,Huang}. Moreover, via constructing a
QHE which is a two-level quantum system and undergoes quantum adiabatic
process and energy exchanges with heat baths at different stages in a work
cycle, some important aspects of the second law of thermodynamics have been
clarified by Kieu \cite{Kieu}. Very recently, many other investigations have
been carried out and demonstrated about the Carnot statement of the second
law of thermodynamics and the quantum Jarzynski equality in quantum systems
described by pseudo-Hermitian Hamiltonians \cite{Gardas}.

While, in contrast to those quantum devices, external reservoirs are
inevitable. QHEs considered in the literature are mainly described by a
Hermitian Hamiltonian and imposed reservoirs. With the proposal and
development of theoretical explorations \cite{Muga,Graefe,Lee},
non-Hermitian quantum theories arise as effective descriptions of certain
open quantum systems, and therefore induce to an effective non-Hermitian QHE
in the presence of absorption and gain. It is natural to establish a
non-Hermitian description for QHEs.

In this paper, we employ a non-Hermitian $\mathcal{PT}$-symmetric
Hamiltonian to study a QHE. It is a two-level system, which describes an $%
S=1/2$ spin in a complex external magnetic field. The non-Hermitian QHE
behaves quite differently from a Hermitian one in all the approaches so far,
providing an alternative description. We do not provide a Hamiltonian to
describe the working medium, whereas a fully description of working medium
and external reservoirs is included. The process of exchanging heat can be
adiabatic in the context of quantum mechanics. We will show that such a
non-Hermitian system can fully describe a QHE without imposed external
reservoirs, which can operate at Otto efficiency for an optimal cycle. In
order to get a clear physical picture of the non-Hermitian engine we
construct a classical analogue of this scheme, which is also a variable-mass
cycle.

\section{Model}

We consider a non-Hermitian spin-$1/2$ system in an external magnetic field,
which can be described by the following Hamiltonian%
\begin{equation}
H=\vec{B}\cdot \vec{\sigma},  \label{H}
\end{equation}%
where $\vec{B}=\left( B_{x},B_{y},i\gamma J,\right) $ is a time-dependent
complex magnetic field and $\vec{\sigma}=\left( \sigma _{x},\sigma
_{y},\sigma _{z}\right) $\ is Pauli matrices. Taking%
\begin{eqnarray}
B_{x} &=&J\cos \phi ,B_{y}=-J\sin \phi ,  \label{field} \\
\cos \theta &=&i\gamma /\sqrt{1-\gamma ^{2}},  \nonumber
\end{eqnarray}%
we rewrite the Hamiltonian as the form%
\begin{equation}
H=J\left( t\right) \sqrt{1-\gamma ^{2}}\left(
\begin{array}{cc}
\cos \theta & \sin \theta e^{i\phi \left( t\right) } \\
\sin \theta e^{-i\phi \left( t\right) } & -\cos \theta%
\end{array}%
\right)  \label{H matrix}
\end{equation}%
where $J\left( t\right) $ and $\phi \left( t\right) $ are real functions of
time, and $\theta (\gamma )$ is a complex (real) constant. Obviously, the
Hamiltonian $H$ is $\mathcal{PT}$-symmetric, i.e., $[\mathcal{P},H]\neq 0$
and $[\mathcal{T},H]\neq 0$, but $[\mathcal{PT},H]=0$, in the sense of $%
\mathcal{T}i\mathcal{T}=-i$ and $\mathcal{P}=\sigma _{x}$. The diagonal
elements of the matrix can be regarded as imaginary on-site potentials of a
two-site tight-binding model, the existence of which violates the law of
conservation of mass and energy. It is thought as an open system with the
source and drain. The aim of this paper is trying to establish a full
quantum mechanical description of the QHE. We will consider a non-Hermitian
two-level quantum engine, in which the source and drain are thought as the
channel to the heat bath. To this end, we will seek the solution of the
Schrodinger equation of the system.
\begin{figure}[tbp]
\begin{center}
\includegraphics[bb=3 424 572 810, width=9cm, clip]{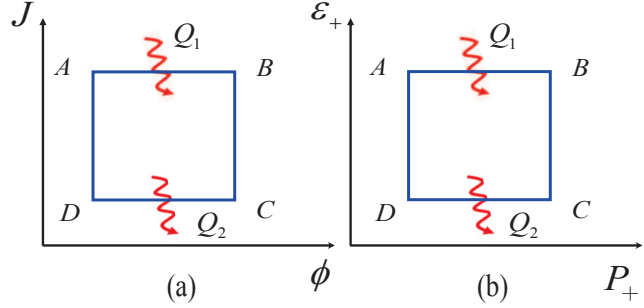}
\end{center}
\caption{(Color online) Schematic diagrams of a variable-mass Otto cycle for
the non-Hermitian QHE. The cycle is shown as a closed loop in the (a) $J-%
\protect\phi $\ and (b) $\protect\varepsilon _{+}-P_{+}$\ planes,
respectively, consisting of four steps, two isospectrum processes $%
A\rightarrow B$ and\ $C\rightarrow D$;\ and two adiabatic processes $%
B\rightarrow C$ and\ $D\rightarrow A$. According to Eq. (\protect\ref{DW}),
the mechanical work done by the engine equals the surface integral of a
nonzero function over the surface enclosed by the loop.}
\label{fig1}
\end{figure}
Considering the adiabatic time-evolution of an initial eigenstate of $%
H\left( 0\right) $\ under the quantum adiabatic condition, we have
\begin{equation}
\left\vert \Psi _{\lambda }\left( t\right) \right\rangle =e^{i\Lambda
_{\lambda }}\left\vert \psi _{\lambda }\left( t\right) \right\rangle ,
\label{solution}
\end{equation}%
where $\left\vert \psi _{\lambda }\left( t \right) \right\rangle $\textbf{%
\ }$\left( \lambda =\pm \right) $\ is the instantaneous eigenstate of $%
H\left( t\right) $,%
\begin{equation}
\left\vert \psi _{+}\right\rangle =\left(
\begin{array}{c}
\cos \frac{\theta }{2} \\
\sin \frac{\theta }{2}e^{-i\phi }%
\end{array}%
\right) ,\left\vert \psi _{-}\right\rangle =\left(
\begin{array}{c}
-\sin \frac{\theta }{2} \\
\cos \frac{\theta }{2}e^{-i\phi }%
\end{array}%
\right) ,  \label{Psi+/-}
\end{equation}%
with eigenvalue $\varepsilon _{\lambda }=\lambda J\left( t\right) \sqrt{%
1-\gamma ^{2}}$ and $J\left( t\right) >0$\ (see Appendix). Phase $\Lambda
_{\lambda }$ is the key quantity of this work, which can be expressed as%
\begin{eqnarray}
\Lambda _{\lambda } &=&-\lambda \sqrt{1-\gamma ^{2}}\int_{0}^{t}J\left(
t\right) \mathrm{d}t  \nonumber \\
&&+\frac{1}{2}\left[ \phi \left( t\right) -\phi \left( 0\right) \right]
\left( 1-i\lambda \gamma /\sqrt{1-\gamma ^{2}}\right) .
\end{eqnarray}%
In the rest of the paper, our investigation does not involve a coherent
superposition of $\left\vert \psi _{\lambda }\left( t \right)
\right\rangle $. We neglect the real part of $\Lambda _{\lambda }$\ and take%
\begin{equation}
\Lambda _{\lambda }=-\frac{i\lambda \gamma }{2}\left[ \phi \left( t\right)
-\phi \left( 0\right) \right] /\sqrt{1-\gamma ^{2}}.  \label{Lambda}
\end{equation}

We note that $\Lambda _{\lambda }$\ is path independent and proportional to
the difference of $\phi $. And the nonzero imaginary adiabatic phase is
quite essential to the thermal process of QHE, since it can vary the
amplitude of instantaneous eigenstates \cite{LS}, leading to the change of
population distributions. The non-Hermitian Hamiltonian is rather different
from a Hermitian one, treating the system as a whole in describing the
working medium and reservoirs.

Before start dealing with a heat engine cycle, we would like to briefly
remark on the time evolution of a non-Hermitian system, which involves both
non-unitary evolution and complex geometrical phase of a state. The Schr\"{o}%
dinger equation is the common basis of both conventional and non-Hermitian
systems. A quantum state evolves along the solution of the Schr\"{o}dinger
equation. According to the solution, the Dirac probability is no longer
conservative. In this sense, the non-unitary evolution and the complex
geometric phase are nothing but the direct and natural results of the Schr%
\"{o}dinger equation.
\begin{table*}[tbp]
\caption{State parameters for a quantum variable-mass Otto cycle illustrated
in Fig \protect\ref{fig1}. Here $T,\protect\varepsilon _{+}$ and $U$ are in
the unit of $\protect\sqrt{1-\protect\gamma ^{2}}$.}
\label{Table I}
\begin{center}
\renewcommand\arraystretch{1}
\par
\begin{tabular}{cccccccc}
\hline\hline
& \parbox{0.5cm} {$J$} & \parbox{0.5cm} {$\phi$} & \parbox{2cm}   {$T$} & %
\parbox{0.5cm} {$\varepsilon _{+}$} & \parbox{2.2cm} {$P_{+},P_{-}$} & %
\parbox{2cm}   {$U$} & \parbox{1cm}   {$S$} \\ \hline
A & \parbox{0.5cm} { $J_{1}$ } & \parbox{0.5cm} {$\phi _{1}$} & %
\parbox{2.2cm} {$\frac{2J_{1}}{k_{\mathrm{B}}\ln\left(p_{0}^{-1}-1\right)}$}
& \parbox{0.5cm}    {$J_{1}$} & \parbox{2.2cm} {$p_{0},1-p_{0}$} & %
\parbox{2cm} {$J_{1}\left( 2p_{0}-1\right)$} &
\parbox{4.5cm}    {$\begin{array}{c}
                                    -k_{\mathrm{B}}p_{0}\ln p_{0}\\
                                    -k_{\mathrm{B}}\left( 1-p_{0}\right)\ln \left(1-p_{0}\right)
                                   \end{array}$} \\
&  &  &  &  &  &  &  \\
B & \parbox{0.5cm} {$J_{1}$} & \parbox{0.5cm}    {$\phi _{2}$} &
\parbox{2.2cm}       {$\frac{2J_{1}}{k_{\mathrm{B}}\ln \left[
\left(p_{0}^{-1}-1\right) \xi ^{-2}\right] }$} & \parbox{0.5cm}
{$J_{1}$} & \parbox{2.2cm}   {$p_{0}\xi ,\left(1-p_{0}\right) \xi ^{-1}$} &
\parbox{2cm}       {$\begin{array}{c}
                                        p_{0}\left( \xi +\xi ^{-1}\right) J_{1}\\    -\xi ^{-1}J_{1}
                                        \end{array}$} &
\parbox{5cm}        {$\begin{array}{c}
                                      -k_{\mathrm{B}}p_{0}\xi \ln \left( p_{0}\xi \right)  \\
                                      -k_{\mathrm{B}}\left( 1-p_{0}\right) \xi ^{-1}\ln \left[ \left(
                                      1-p_{0}\right) \xi ^{-1}\right]
                                        \end{array}$} \\
&  &  &  &  &  &  &  \\
C & \parbox{0.5cm} {$J_{2}$} & \parbox{0.5cm} {$\phi _{2}$} &
\parbox{2.2cm}
{$\frac{2J_{2}}{k_{\mathrm{B}}\ln \left[ \left(p_{0}^{-1}-1\right) \xi
^{-2}\right] }$} & \parbox{0.5cm}    {$J_{2}$} &
\parbox{2.2cm}   {$p_{0}\xi
,\left(1-p_{0}\right) \xi ^{-1}$} &
\parbox{2cm}   {$\begin{array}{c}
                                   p_{0}\left( \xi +\xi ^{-1}\right) J_{2}\\  -\xi ^{-1}J_{2}
                                    \end{array}$} &
\parbox{5cm}   {$\begin{array}{c}
                                 -k_{\mathrm{B}}p_{0}\xi \ln \left( p_{0}\xi \right)  \\
                                 -k_{\mathrm{B}}\left( 1-p_{0}\right) \xi ^{-1}\ln \left[ \left( 1-p_{0}\right) \xi ^{-1}\right]
                                \end{array}$} \\
&  &  &  &  &  &  &  \\
D & \parbox{0.5cm} {$J_{2}$} & \parbox{0.5cm}   {$\phi _{1}$} &
\parbox{2.2cm}
{$\frac{2J_{2}}{k_{\mathrm{B}}\ln\left(p_{0}^{-1}-1\right)}$} & %
\parbox{0.5cm}      {$J_{2}$} & \parbox{2.2cm}   {$p_{0},1-p_{0}$} & %
\parbox{2cm}   {$J_{2}\left( 2p_{0}-1\right)$} &
\parbox{4.5cm}      {$\begin{array}{c}
                                      -k_{\mathrm{B}}p_{0}\ln p_{0}\\
                                      -k_{\mathrm{B}}\left( 1-p_{0}\right)\ln \left(1-p_{0}\right) \end{array}$}
\\ \hline\hline
\end{tabular}%
\end{center}
\end{table*}
\begin{table*}[tbp]
\caption{Variances of state parameters related to four processes in a
quantum variable-mass Otto cycle illustrated in Fig \protect\ref{fig1}. Here
$\Delta T$ and $\Delta U$ are in the unit of $\protect\sqrt{1-\protect\gamma %
^{2}}$.}
\label{Table II}
\begin{center}
\begin{tabular}{cccccc}
\hline\hline
\parbox{1.2cm} {$ $} & \parbox{2cm} {$\Delta T$} & \parbox{3cm} {$\Delta U$}
& \parbox{0.5cm} {$\Delta Q$} & \parbox{0.5cm} {$\Delta W$} &
\parbox{0.5cm}
{$\Delta S$} \\ \hline
\parbox{1.2cm} {$A \rightarrow B$} &
\parbox{3.8cm}    {$\frac{4J_{1}\ln \xi
}{k_{\mathrm{B}}\ln \left[ \left(p_{0}^{-1}-1\right) \xi ^{-2}\right] \ln
\left( p_{0}^{-1}-1\right) }$ } &
\parbox{3.5cm}    {$\begin{array}{c}
                                    p_{0}\left( \xi +\xi ^{-1}-2\right) J_{1} \\
                                   +\left( 1-\xi ^{-1}\right) J_{1}
                                     \end{array}$} &
\parbox{0.5cm}
{$\Delta U$} & \parbox{0.5cm} {0} &
\parbox{4.7cm}   {$\begin{array}{c}
                                   -k_{\mathrm{B}}p_{0}\ln \left( p_{0}^{\xi -1}\xi ^{\xi }\right)-k_{\mathrm{B}}\left( 1-p_{0}\right)  \\
                                  \times\ln \left[ \left( 1-p_{0}\right) ^{\xi^{-1}-1}\left( \xi ^{-1}\right) ^{\xi ^{-1}}\right]
                                  \end{array}$} \\
&  &  &  &  &  \\
\parbox{1.2cm} {$B \rightarrow C$} &
\parbox{3.8cm} {
$\frac{2\left(J_{2}-J_{1}\right) }{k_{\mathrm{B}}\ln \left[
\left(p_{0}^{-1}-1\right) \xi ^{-2}\right] }$ } &
\parbox{3.3cm}    {$\begin{array}{c}
                                      p_{0}\left( \xi +\xi ^{-1}\right) \left( J_{2}-J_{1}\right)  \\
                                      -\xi ^{-1}\left( J_{2}-J_{1}\right)
                                      \end{array}$} & \parbox{0.5cm}    {0}
& \parbox{0.5cm}    {$\Delta U$} & \parbox{0.8cm}    {0} \\
&  &  &  &  &  \\
\parbox{1.2cm}   {$C \rightarrow D$} &
\parbox{3.8cm}   {$-\frac{4J_{2}\ln
\xi }{k_{\mathrm{B}}\ln \left[ \left(p_{0}^{-1}-1\right) \xi ^{-2}\right]
\ln \left(p_{0}^{-1}-1\right) }$ } &
\parbox{3.5cm} {$\begin{array}{c}
                                  -p_{0}\left( \xi +\xi ^{-1}-2\right) J_{2} \\
                                  -\left( 1-\xi ^{-1}\right) J_{2}
                                  \end{array}$} &
\parbox{0.5cm}   {$\Delta
U$} & \parbox{0.5cm} {0} &
\parbox{4.7cm}   {$\begin{array}{c}
                                    k_{\mathrm{B}}p_{0}\ln \left( p_{0}^{\xi -1}\xi ^{\xi }\right)+k_{\mathrm{B}}\left( 1-p_{0}\right)  \\
                                   \times\ln \left[ \left( 1-p_{0}\right) ^{\xi
                                   ^{-1}-1}\left( \xi ^{-1}\right) ^{\xi ^{-1}}\right]
                                  \end{array}$} \\
&  &  &  &  &  \\
\parbox{1.2cm} {$D \rightarrow A$} &
\parbox{3.8cm} {
$\frac{2\left(J_{1}-J_{2}\right) }{k_{\mathrm{B}}\ln
\left(p_{0}^{-1}-1\right) }$ } &
\parbox{3.3cm}
{$\left(J_{1}-J_{2}\right)\left( 2p_{0}-1\right)$} & \parbox{0.5cm} {0} & %
\parbox{0.5cm} {$\Delta U$} & \parbox{0.8cm} {0} \\ \hline\hline
\end{tabular}%
\end{center}
\end{table*}

\section{All-quantum-adiabatic cycle}

Now we investigate a thermodynamic process for an adiabatic passage of the
non-Hermitian system.\textbf{\ }For a traditional (Hermitian) QHE, the
temperature of a quantum working medium is introduced based on the
thermalization assumption \cite{Quan2007,Henrich,Kieu,Gemmer}, that is, the
working medium has the same distribution of occupation probability as the
heat bath in thermodynamic equilibrium. Then for such a small quantum
system, the probability distribution is related to a parameter called
spectral temperature. In this paper, we still employ such a concept for the
non-Hermitian two-level system, as an assumption. Consider an initially
prepared mixed state, which has Dirac probability $P_{\lambda }(t=0)$\ on
the eigenstate $\left\vert \psi _{\lambda }\left( t=0\right) \right\rangle $
 of $H\left( 0\right) $\ with eigenenergy%
\textbf{\ }$\lambda J_{0}\sqrt{1-\gamma ^{2}}$. The system parameters are
taken as $\left( J_{0},\phi _{0}\right) $\ initially where $J_{0}=J\left(
0\right) $ and $\phi _{0}=\phi \left( 0\right) $. And the initial
populations for the upper and lower levels are $P_{+}(0)=p_{0}$\ and $%
P_{-}(0)=1-p_{0}$, respectively. We express this mixed state by a density
matrix%
\begin{equation}
\rho (0)=\sum_{\lambda =\pm }P_{\lambda }(0)\left\vert \psi _{\lambda
}\left( 0\right) \right\rangle \left\langle \psi _{\lambda }\left( 0\right)
\right\vert ,
\end{equation}%
where eigenstate $\left\vert \psi _{\lambda }\left( 0\right) \right\rangle $%
\ is normalized in the context of Dirac inner product (see Appendix). Here
we assume that the Boltzmann distribution\ still refers to Dirac probability
even for a non-Hermitian system. When a two-level system\ couples to a heat
bath, both systems obey the thermal equilibrium Boltzmann distributions and
then the two-level mixed state has its temperature.\ A density matrix would
represent a thermal equilibrium state with temperature

\begin{equation}
T_{0}=\frac{2J_{0}\sqrt{1-\gamma ^{2}}}{k_{\mathrm{B}}\ln \left(
p_{0}^{-1}-1\right) },  \label{T0}
\end{equation}%
where $k_{\mathrm{B}}$\ is the Boltzmann constant. In quantum statistics,
the definition of temperature does not require the preservation of particle
probability. It is only determined by the distribution of particle
probability on each level. For a two-level system, one can always find a
temperature which only depends on the ratio of the probabilities and the energy
difference of two levels, i.e.,
\begin{equation}
\ln \frac{P_{+}}{P_{-}}=-\frac{\varepsilon _{+}-\varepsilon _{-}}{k_{\mathrm{%
B}}T}
\end{equation}%
where $P_{\lambda}$ and $\varepsilon _{\lambda} $ are the particle
probability and the energy of two levels. At time $t$, state $\left\vert
\psi _{\lambda }\left( 0\right) \right\rangle $ evolves into state $%
\left\vert \Psi _{\lambda }\left( t\right) \right\rangle $\ under the
time-dependent Hamiltonian $H\left( t\right) $. We note that the initial
state does not involve a coherent superposition of $\left\vert \psi
_{+}\left( 0\right) \right\rangle $ and $\left\vert \psi _{-}\left( 0\right)
\right\rangle $. Then the evolved mixed state does not involve a coherent
superposition of $\left\vert \Psi _{+}\left( t\right) \right\rangle $ and $%
\left\vert \Psi _{-}\left( t\right) \right\rangle $. We only concern about
the time evolution of the pure state $\left\vert \Psi _{\lambda }\left(
t\right) \right\rangle $.

It is an acceptable consensus that the Schr\"{o}dinger equation is the
common basis for both Hermitian and non-Hermitian quantum mechanics. Whether
the operator $H\left( t\right) $\ is Hermitian or not, an evolved state,
vector $\left\vert \Psi _{\lambda }\left( t\right) \right\rangle $ should be
the solution of the Schr\"{o}dinger equation%
\begin{equation}
i\frac{\partial }{\partial t}\left\vert \Psi _{\lambda }\left( t\right)
\right\rangle =H\left( t\right) \left\vert \Psi _{\lambda }\left( t\right)
\right\rangle ,
\end{equation}%
with the initial condition $\left\vert \Psi _{\lambda }\left( 0\right)
\right\rangle =\left\vert \psi _{\lambda }\left( 0\right) \right\rangle $.
Accordingly, the density matrix of the evolved mixed state becomes

\begin{equation}
\rho (t)=\sum_{\lambda =\pm }P_{\lambda }(0)\left\vert \Psi _{\lambda
}\left( t\right) \right\rangle \left\langle \Psi _{\lambda }\left( t\right)
\right\vert ,
\end{equation}%
where $\left\vert \Psi _{\lambda }\left( t\right) \right\rangle $\ is
probably not Dirac normalized so as to contribute to the population $%
P_{\lambda }(t)$. However, it is tough to get an analytical solution for an arbitrary time-dependent operator $H\left( t\right) $. Fortunately, when we
only consider the time evolution under the quantum adiabatic condition,
i.e., there is no transition between two levels, an approximate solution can
be obtained. Then the density matrix of the evolved mixed state can be
expressed as

\begin{equation}
\rho (t)=\sum_{\lambda =\pm }P_{\lambda }(t)\left\vert \psi _{\lambda
}\left( t\right) \right\rangle \left\langle \psi _{\lambda }\left( t\right)
\right\vert ,
\end{equation}%
where%
\begin{equation}
P_{\lambda }(t)=P_{\lambda }(0)\exp [-2\textrm{Im}(\Lambda _{\lambda })].
\end{equation}%
And the corresponding temperature of the thermal state $\rho (t)$ is

\begin{equation}
T(t)=\frac{2J(t)\sqrt{1-\gamma ^{2}}}{k_{\mathrm{B}}\ln \left[ \left(
p_{0}^{-1}-1\right) \xi ^{-2}\right] },
\end{equation}%
where $\xi =\exp \{\gamma \left[ \phi (t)-\phi _{0}\right] /\sqrt{1-\gamma
^{2}}\}$ characterizes the amplification or attenuation of the whole
probability in the working medium.

Under the quantum adiabatic condition, the solution of Eq. (\ref{solution})
tells us that any evolution along an arbitrary path in the $J-\phi $ plane
is always adiabatic in the context of quantum mechanics, i.e., there is no
transition between two levels. Nevertheless, it does not mean that the
occupation probabilities for the levels and their distributions are always
remaining invariant due to the contribution of the $\Lambda _{\lambda }$ in
Eq. (\ref{Lambda}), which are unusual and exotic properties in the
non-Hermitian system in contrast to the Hermitian one. Moreover, we would
demonstrate that the \textquotedblleft adiabatic\textquotedblright\ in our
paper for the non-Hermitian Hamiltonian may be different from the one in the
Hermitian system because the occupation probability for each level is not
invariant due to the image part of $\Lambda _{\lambda }$. But what both
adiabatic processes of the Hermitian and non-Hermitian system have in common
is that there is no tunneling between each level. Therefore here we could
call it \textquotedblleft generalized adiabatic\textquotedblright\ in our
non-Hermitian system. Since all the processes considered in our work are
\textquotedblleft generalized adiabatic\textquotedblright , we only write it
as the \textquotedblleft adiabatic\textquotedblright\ for simplification.
Now we consider a loop in the $J-\phi $ plane as a thermal cycle.

According to the first law of thermodynamics, the total heat transferred in
the cycle is
\begin{eqnarray}
\Delta Q &=&\oint_{\mathrm{c}}\mathrm{d}Q=\oint_{\mathrm{c}}\sum_{\lambda
}\varepsilon _{\lambda }\mathrm{d}P_{\lambda }  \nonumber \\
&=&\frac{2\gamma }{Z}\int \!\!\!\int_{\mathrm{A}}\cosh \left( \ln \sqrt{%
\frac{1}{p_{0}}-1}-\ln \xi \right) \mathrm{d}J\mathrm{d}\phi ,  \label{DQ}
\end{eqnarray}%
where $Z=2\cosh (\ln \sqrt{\frac{1}{p_{0}}-1})$ is the partition function. Here $\mathrm{A}$\ denotes the area
enclosed by the loop \textrm{c} in the $J-\phi $ plane. We can see that $%
\Delta Q$ is nonzero for a non-trivial cycle, since a hyperbolic cosine
function is positive definite. Similarly, the net work done in the cycle is
\
\begin{equation}
\Delta W=\oint_{\mathrm{c}}\mathrm{d}W=\oint_{\mathrm{c}}\sum_{\lambda
}P_{\lambda }\mathrm{d}\varepsilon _{\lambda }=-\Delta Q,  \label{DW}
\end{equation}%
which leads to
\begin{equation}
\Delta U=\Delta Q+\Delta W=0.  \label{dU=0}
\end{equation}%
Furthermore, a straightforward derivation shows that the entropy variation
in the loop is%
\begin{equation}
\Delta S=\oint_{\mathrm{c}}\frac{1}{T}\mathrm{d}Q=\oint_{\mathrm{c}}\frac{1}{%
T}\sum_{\lambda }\varepsilon _{\lambda }\mathrm{d}P_{\lambda }=0.  \label{DS}
\end{equation}

According with the traditional thermodynamics, these indicate that the
internal energy $U$\ and entropy $S$\ are state functions, which are only
determined by the system parameters $\left( J,\phi \right) $. We would like
to point out that the facts of $\Delta Q\neq 0$ and $\Delta W\neq 0$\ in the
loop are exclusive for a non-Hermitian system. Taking $\gamma =0$, the
system is reduced to a Hermitian one. We can see that $\Delta Q=$ $\Delta
W=0 $, which indicates that the Hermitian system is trivial as a QHE. More
detailed, as mentioned above, the complex parameter represents the effect of
reservoirs phenomenologically. Once $\gamma =0$, the working medium is
decoupled from reservoirs. There is no exchange between the working medium
and the reservoirs. Then the heat engine becomes trivial, since we do not
assume extra reservoirs imposed on the system. This is different from the
conventional Hermitian QHE, which is always accompanied by the reservoirs.
This reflects the key point of our work, replacing the imposed reservoirs by
the imaginary parameter $\gamma $. In the context of quantum mechanics,
processes for an arbitrary loop is spontaneously adiabatic without any extra
condition. This may enhance the feasibility of experimental realizations of
the present scheme.

\section{Cycle with Otto efficiency}

So far we have established an alternative description for QHEs via a
non-Hermitian Hamiltonian. Remarkably, all the thermal processes are
adiabatic in a quantum mechanics manner. By this we mean that irrespective
of whether there is energy transfer between the heat engine and reservoirs
or not all the processes are adiabatic evolutions driven by a time-dependent
Hamiltonian. To understand such quantum processes, it is useful to contrast
them to classical thermal processes. In the following, we try to connect
such a description to a classical engine, showing that the non-Hermitian QHE
is not a non-physics toy model.

To this end, we consider a specific cycle, which consists of four processes
connecting four points, $A$: $\left( J_{1},\phi _{1}\right) $, $B$: $\left(
J_{1},\phi _{2}\right) $, $C$: $\left( J_{2},\phi _{2}\right) $, and $D$: $%
\left( J_{2},\phi _{1}\right) $. There are two types of processes involved:
(I) Isospectrum process, $A\rightarrow B$ and\ $C\rightarrow D$; (II)
Adiabatic process, $B\rightarrow C$ and\ $D\rightarrow A$.\ A schematic
illustration is listed in Figs. \ref{fig1}(a) and \ref{fig1}(b). In type I
process, varying $\phi $\ can change the populations on two levels due to
the imaginary adiabatic phase $\Lambda _{\lambda }$, but leave the level
structure unchanged. The level structure can be completely characterized by
quantity $\varepsilon _{+}$. While, in type II process, varying $J$\ can
change the level structure $\varepsilon _{+}$, whereas leave populations on
two levels unchanged. Based on the adiabatic solution of Eq. (\ref{solution}%
), quantities, including the state parameters at each point and their
variances in each process, are obtained and listed in Tables \ref{Table I}\
and \ref{Table II}, respectively. According to the definition in the Refs.
\cite{Kieu,Quan2007,Henrich}, our calculations indicate that such an engine
can operate at Otto efficiency%
\begin{equation}
\eta =\frac{Q_{\mathrm{1}}+Q_{\mathrm{2}}}{Q_{\mathrm{2}}}=1-\frac{J_{2}}{%
J_{1}}.  \label{efficiency}
\end{equation}

Such a non-Hermitian QHE does not violate the Carnot efficiency limit which
is of great significance in thermodynamics. This indicates that the
non-Hermitian Hamiltonian still obeys some underlying rules in nature.

We note that the probability is not conservative in the cycle, which differs
from that in a Hermitian system. We refer to this cycle as a variable-mass
Otto cycle. However, we would like to point out that the particle-number
conservation still holds in such a cycle \cite{JL2}. Here we use the term
variable-mass in order to correspond with its classical analogue. For a
particle-exchange heat engine, it has been investigated in Ref. \cite%
{Humphrey}. Moreover, to understand such a non-Hermitian cycle and its
related issues, we propose a variable-mass cycle for classical ideal gas in
the following.

\begin{figure}[tbp]
\begin{center}
\includegraphics[bb=3 424 572 810, width=9cm, clip]{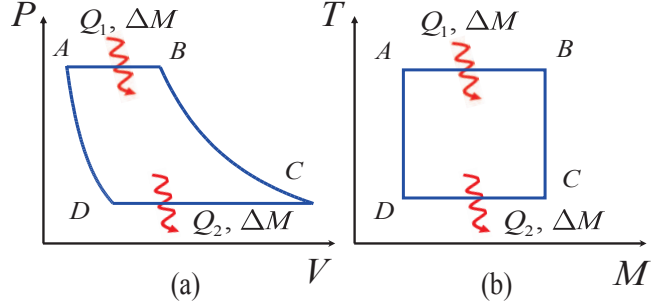}
\end{center}
\caption{(Color online) A classical analogue to the quantum variable-mass
cycle consisting of four steps, two isothermal processes and two adiabatic
processes. Here the isothermal process is associated with particle transfer.
The cycle is shown as a closed loop in the\ (a) $P-V$\ and (b) $T-M$\
planes, respectively. }
\label{fig2}
\end{figure}

\section{Classical analogue}

In order to obtain a clear physical picture of the non-Hermitian QHE we
construct a classical analogue of this scheme, which is a variable-mass
cycle and consists of the following four steps, which is schematically
illustrated in Fig. \ref{fig2}.

1. Reversible isothermal addition of gas at the high temperature $T_{1}$.
During this step a volume $\Delta V_{1}$ of the gas with the same
temperature $T_{1}$\ and the same density as the working medium is added. It
does no work on the surroundings. The temperature of the gas does not change
during the process, and thus the addition is isothermal. The gas addition is
associated by absorption of heat energy $Q_{1}$ and increase of entropy $%
\Delta S=Q_{1}/T_{1}$ from the high temperature reservoir.

2. Reversible adiabatic expansion of gas. For this step, the gas is allowed
to expand and does work on the surroundings, whose expansion causes it to
cool to the low temperature $T_{2}$, with the entropy remaining unchanged.

3. Reversible isothermal deduction of gas at the low temperature $T_{2}$.
During this step a volume $\Delta V_{2}$ of the gas with the same
temperature $T_{2}$\ and the same density as the working medium is removed.
Such an amount of the gas contains heat energy $Q_{2}$ and entropy $\Delta
S=Q_{2}/T_{2}$, which flows to the low temperature reservoir.

4. Reversible adiabatic compression of gas. During this step, the
surroundings do work on the gas, increasing its internal energy and
compressing it, causing the temperature to rise to $T_{1}$. The gas goes
back to the state at the start.

For a standard cycle, mass is conserved, i.e., the working medium does not
lose or increase mass. The present cycle transfers mass from hot to cool
reservoirs, associating with the heat transfer. The essence of this cycle is
that there is a volume of gas transferred from hot to cool reservoirs,
rather than transfer the heat only. Then the heat efficiency is always $%
1-T_{2}/T_{1}$, which is independent of the mass transferred. Therefore, the
proposed variable-mass cycle would help us to make a better understanding of
the non-Hermitian cycle in some sense.

\section{Summary and discussion}

We have studied a time-dependent non-Hermitian $\mathcal{PT}$-symmetric
two-level system, which could be regarded as a QHE. A non-Hermitian
Hamiltonian turns out to be quite different from a Hermitian one. It does
not solely describe the working medium, but offers a full description
including the working medium and external reservoirs. We have shown that any
cycle can be implemented by an adiabatic time-evolution along a quantum
adiabatic passage. As an example, a specific cycle which operates at Otto
efficiency is presented. Moreover, we have also discussed a classical
analogue of this scheme, which corresponds to a variable-mass cycle,
transferring not only heat but also mass from hot to cool reservoirs. Our
main conclusion is that the proposed non-Hermitian Hamiltonian can provide
an alternative description for the QHE, leading to a deeper and better
understanding of the role that a non-Hermitian Hamiltonian takes in physics.

Finally, we would like to point that there is a long way to go for
connecting the current non-Hermitian 2-level system to a Hermitian
description of QHE by using a rigorous mathematical formulation. It is due
to the subtle relation between Hermitian and non-Hermitian Hamiltonians: on
the one hand, a non-Hermitian Hamiltonian is usually thought to be a better
candidate for describing some natural processes in an open system, which do
not follow the conservation of law of mass and energy. However, on the other
hand, it is still a challenge to get a perfect connection between a
non-Hermitian Hamiltonian and a Hermitian one with a clear physical picture,
although many efforts have been dedicated to this topic \cite{JL1}.

\section{Appendix}

The instantaneous eigenstates of $H\left( t\right) $ and $H^{\dagger }\left(
t\right) $\ with eigenvalue $\varepsilon _{\lambda }=\lambda J\left(
t\right) \sqrt{1-\gamma ^{2}}$ are%
\begin{equation}
\left\vert \psi _{+}\right\rangle =\left(
\begin{array}{c}
\cos \frac{\theta }{2} \\
\sin \frac{\theta }{2}e^{-i\phi }%
\end{array}%
\right) ,\left\vert \psi _{-}\right\rangle =\left(
\begin{array}{c}
-\sin \frac{\theta }{2} \\
\cos \frac{\theta }{2}e^{-i\phi }%
\end{array}%
\right) ,
\end{equation}%
and%
\begin{equation}
\left\vert \eta _{+}\right\rangle =\left(
\begin{array}{c}
\cos \frac{\theta }{2} \\
\sin \frac{\theta }{2}e^{i\phi }%
\end{array}%
\right) ^{\ast },\left\vert \eta _{-}\right\rangle =\left(
\begin{array}{c}
-\sin \frac{\theta }{2} \\
\cos \frac{\theta }{2}e^{i\phi }%
\end{array}%
\right) ^{\ast },
\end{equation}%
respectively, which can construct biorthonormal complete set $\left\{
\left\vert \psi _{\lambda }\right\rangle ,\left\vert \eta _{\lambda
}\right\rangle \right\} $ ($\lambda =\pm $), i.e.,

\begin{equation}
\left\langle \eta _{\lambda ^{\prime }}\right. \left\vert \psi _{\lambda
}\right\rangle =\delta _{\lambda \lambda ^{\prime }},\sum_{\lambda
}\left\vert \psi _{\lambda }\right\rangle \left\langle \eta _{\lambda
}\right\vert =1.
\end{equation}%
Here we demonstrate that when computing the density matrix or detecting the
probability in experiments, both eigenstates of $H\left( t\right) $\ must be
normalized in Dirac inner product, i.e., we need to take $\left\vert \psi
_{\lambda }\right\rangle \rightarrow \left\vert \psi _{\lambda
}\right\rangle /\sqrt{\Theta }$\ with $\Theta =\left\vert \cos \frac{\theta
}{2}\right\vert ^{2}+\left\vert \sin \frac{\theta }{2}\right\vert ^{2}$. Now
we extend the adiabatic theorem to the non-Hermitian Hamiltonian and we will
show that such an extension is correct. This extension can be simply applied
by taking the biorthonormal inner product to replace the Dirac inner
product. Then under the quantum adiabatic condition
\begin{equation}
\left\vert \frac{\left\langle \eta _{-}\right\vert \frac{\partial H}{%
\partial t}\left\vert \psi _{+}\right\rangle }{\left( \varepsilon
_{+}-\varepsilon _{-}\right) ^{2}}\right\vert =\left\vert \frac{\dot{\phi}}{%
4J\left( t\right) \left( 1-\gamma ^{2}\right) }\right\vert \ll 1,
\label{ad cond}
\end{equation}%
phase $\Lambda _{\lambda }$ can be expressed as%
\begin{eqnarray}
\Lambda _{\lambda } &=&-\int_{0}^{t}\varepsilon _{\lambda }\mathrm{d}%
t+i\int_{0}^{t}\left\langle \eta _{\lambda }\right\vert \frac{\partial }{%
\partial t}\left\vert \psi _{\lambda }\right\rangle \mathrm{d}t  \nonumber \\
&=&-\lambda \sqrt{1-\gamma ^{2}}\int_{0}^{t}J\left( t\right) \mathrm{d}t
\nonumber \\
&&+\frac{1}{2}\left[ \phi \left( t\right) -\phi_0 \right]
\left( 1-i\lambda \gamma /\sqrt{1-\gamma ^{2}}\right) .
\end{eqnarray}%
Similar works have been done in many systems \cite%
{Gao,Nesterov1,Nesterov2,Cui,Liang}, and this conclusion can also be
obtained by an exact solution of Schr\"{o}dinger equation%
\begin{equation}
i\frac{\partial }{\partial t}\left\vert \Psi \right\rangle =H\left\vert \Psi
\right\rangle ,  \label{S.eq}
\end{equation}%
in this concrete example. Here our $\mathcal{PT}$-symmetric Hamiltonian $H$
could be used to describe an open system, which is different from \cite{Gong}
that is used to describe a closed system. Actually, the propagator of the
Hamiltonian $H$%
\begin{equation}
U=\mathcal{T}\exp (-i\int_{0}^{t}H\left( t\right) \mathrm{d}t)
\end{equation}%
can be obtained as the form%
\begin{eqnarray}
U_{11} &=&\left[ \cos \left( \Omega \tau \right) -i\frac{\Delta }{2\Omega }%
\sin \left( \Omega \tau \right) \right] e^{i\omega \tau /2},  \nonumber \\
U_{22} &=&\left[ \cos \left( \Omega \tau \right) +i\frac{\Delta }{2\Omega }%
\sin \left( \Omega \tau \right) \right] e^{-i\omega \tau /2},  \nonumber \\
U_{12} &=&-\frac{i}{\Omega }\sin \left( \Omega \tau \right) e^{i\omega \tau
/2},  \nonumber \\
U_{21} &=&-\frac{i}{\Omega }\sin \left( \Omega \tau \right) e^{-i\omega \tau
/2},
\end{eqnarray}%
when we take $\phi \left( \tau \right) =\omega \tau $. Here $\mathcal{T}$\
is time-order operator and%
\begin{eqnarray}
\tau &=&\int_{0}^{t}J\left( t\right) \mathrm{d}t,\Omega =\frac{1}{2}\sqrt{%
4+\Delta ^{2}}, \\
\Delta &=&2i\gamma +\omega .
\end{eqnarray}%
The adiabatic condition in Eq. (\ref{ad cond}) reduces to
\begin{equation}
\left\vert \frac{\omega }{1-\gamma ^{2}}\right\vert \ll 1,
\end{equation}%
under which the Taylor expansion\ gives

\begin{eqnarray}
-\Omega \tau &=&-\sqrt{1-\gamma ^{2}}\sqrt{1+\frac{i\gamma \omega }{1-\gamma
^{2}}}\tau  \nonumber \\
&\approx &-\int_{0}^{t}\sqrt{1-\gamma ^{2}}J\left( t\right) \mathrm{d}t-%
\frac{i\gamma \omega \tau }{2\sqrt{1-\gamma ^{2}}}.
\end{eqnarray}
This solution indicates that the adiabatic process exists under a certain
condition, i.e., there is no tunnelling between instantaneous eigenstates
during the time evolution. Then the corresponding geometrical phase can be
obtained naturally. While, in the article \cite{Milburn}, dynamics near
exceptional point is considered and it is natural that the adiabatic theorem
cannot hold in that situation. The process cannot hold when the system
closes to the exceptional point (or quantum phase transition point for
breaking $\mathcal{PT}$ symmetry). Actually, similar behavior can occur in a
Hermitian system, for example, an adiabatic process cannot be achieved near
a quantum phase transition point. A simple demonstration is Landau-Zener
formula.

Moreover, in the biorthonormal formalism, compared with right ket $%
\left\vert \Psi \right\rangle $ which obeys the Eq. (\ref{S.eq}), the left
bra $\left\langle \Phi \right\vert $ follows the different evolution
equation
\begin{equation}
i\frac{\partial }{\partial t}\left\vert \Phi \right\rangle =H^{\dag
}\left\vert \Phi \right\rangle ,
\end{equation}%
and correspondingly the propagator
\begin{equation}
\tilde{U}=\mathcal{T}\exp (-i\int_{0}^{t}H^{\dag }\left( t\right) \mathrm{d}%
t)
\end{equation}%
can be obtained as the form%
\begin{eqnarray}
\tilde{U}_{11} &=&\left[ \cos \left( \Omega ^{\ast }\tau \right) -i\frac{%
\Delta ^{\ast }}{2\Omega ^{\ast }}\sin \left( \Omega ^{\ast }\tau \right) %
\right] e^{i\omega \tau /2},  \nonumber \\
\tilde{U}_{22} &=&\left[ \cos \left( \Omega ^{\ast }\tau \right) +i\frac{%
\Delta ^{\ast }}{2\Omega ^{\ast }}\sin \left( \Omega ^{\ast }\tau \right) %
\right] e^{-i\omega \tau /2},  \nonumber \\
\tilde{U}_{12} &=&-\frac{i}{\Omega ^{\ast }}\sin \left( \Omega ^{\ast }\tau
\right) e^{i\omega \tau /2},  \nonumber \\
\tilde{U}_{21} &=&-\frac{i}{\Omega ^{\ast }}\sin \left( \Omega ^{\ast }\tau
\right) e^{-i\omega \tau /2},
\end{eqnarray}%
then we can prove%
\begin{equation}
\tilde{U}^{\dag }U=1.
\end{equation}%
As a result, it is unitary according to the biorthonormal inner product.
However, we would like to point out that the complex phase for a
non-Hermitian system is a natural result (see references \cite%
{Gao,Nesterov1,Nesterov2,Cui,Liang}). The Berry phase and the unitarity of
time evolution are compatible with each other.

Next, explicit derivations of Eq. (\ref{DQ}), (\ref{DW}), and (\ref{DS}) are
given. Together with $P_{+}=p_{0}\xi $, $P_{-}=\left( 1-p_{0}\right) \xi
^{-1}$ (here $\xi $ is only the function of $\phi $) and $\varepsilon
_{\lambda }$, we get the heat exchange, the work done, and the variation of
entropy of any arbitrary path in the $J-\phi $ plane as following,

\begin{eqnarray}
\Delta Q &=&\oint_{\mathrm{c}}\mathrm{d}Q=\oint_{\mathrm{c}}\sum_{\lambda
}\varepsilon _{\lambda }\mathrm{d}P_{\lambda }  \nonumber \\
&=&\frac{\sqrt{1-\gamma ^{2}}}{Z}\oint_{\mathrm{c}}J\mathrm{d}\left[
e^{-\left( \ln \sqrt{\frac{1}{p_{0}}-1}-\ln \xi \right) }-e^{\left( \ln
\sqrt{\frac{1}{p_{0}}-1}-\ln \xi \right) }\right]  \nonumber \\
&=&-\frac{2\sqrt{1-\gamma ^{2}}}{Z}\oint_{\mathrm{c}}J\mathrm{d}\left[ \sinh
\left( \ln \sqrt{\frac{1}{p_{0}}-1}-\ln \xi \right) \right]  \nonumber \\
&=&\frac{2\gamma }{Z}\oint_{\mathrm{c}}J\cosh \left( \ln \sqrt{\frac{1}{p_{0}%
}-1}-\ln \xi \right) \mathrm{d}\phi  \nonumber \\
&=&\frac{2\gamma }{Z}\int \!\!\!\int_{\mathrm{A}}\cosh \left( \ln \sqrt{%
\frac{1}{p_{0}}-1}-\ln \xi \right) \mathrm{d}J\mathrm{d}\phi ,
\end{eqnarray}%
and%
\begin{eqnarray}
\Delta W &=&\oint_{\mathrm{c}}\mathrm{d}W=\oint_{\mathrm{c}}\sum_{\lambda
}P_{\lambda }\mathrm{d}\varepsilon _{\lambda }  \nonumber \\
&=&\frac{\sqrt{1-\gamma ^{2}}}{Z}\oint_{\mathrm{c}}\sum_{\lambda }\left[
\lambda e^{-\lambda \left( \ln \sqrt{\frac{1}{p_{0}}-1}-\ln \xi \right) }%
\right] \mathrm{d}J  \nonumber \\
&=&\frac{\sqrt{1-\gamma ^{2}}}{Z}\oint_{\mathrm{c}}\left[ e^{-\left( \ln
\sqrt{\frac{1}{p_{0}}-1}-\ln \xi \right) }-e^{\left( \ln \sqrt{\frac{1}{p_{0}%
}-1}-\ln \xi \right) }\right] \mathrm{d}J  \nonumber \\
&=&-\frac{2\sqrt{1-\gamma ^{2}}}{Z}\oint_{\mathrm{c}}\sinh \left( \ln \sqrt{%
\frac{1}{p_{0}}-1}-\ln \xi \right) \mathrm{d}J  \nonumber \\
&=&-\frac{2\gamma }{Z}\int \!\!\!\int_{\mathrm{A}}\cosh \left( \ln \sqrt{%
\frac{1}{p_{0}}-1}-\ln \xi \right) \mathrm{d}J\mathrm{d}\phi .
\end{eqnarray}%
Obviously we have
\begin{equation}
\Delta Q+\Delta W=0.
\end{equation}%
Similarly, the variation of entropy is%
\begin{eqnarray}
\Delta S &=&\oint_{\mathrm{c}}\frac{1}{T}\mathrm{d}Q=\oint_{\mathrm{c}}\frac{%
1}{T}\sum_{\lambda }\varepsilon _{\lambda }\mathrm{d}P_{\lambda }  \nonumber
\\
&=&-\frac{2k_{\mathrm{B}}}{Z}\oint_{\mathrm{c}}\left( \ln \sqrt{\frac{1}{%
p_{0}}-1}-\ln \xi \right) \mathrm{d}\left[ \sinh \left( \ln \sqrt{\frac{1}{%
p_{0}}-1}-\ln \xi \right) \right]  \nonumber \\
&=&\frac{2k_{\mathrm{B}}\gamma }{Z\sqrt{1-\gamma ^{2}}}\oint_{\mathrm{c}%
}\left( \ln \sqrt{\frac{1}{p_{0}}-1}-\ln \xi \right) \cosh \left( \ln \sqrt{%
\frac{1}{p_{0}}-1}-\ln \xi \right) \mathrm{d}\phi  \nonumber \\
&=&0.
\end{eqnarray}

\section*{Acknowledgement}

We acknowledge the support of the National Basic Research Program (973
Program) of China under Grant No. 2012CB921900 and the CNSF (Grant No.
11374163).

\end{document}